\documentclass[aps,twocolumn,showpacs,floatfix]{revtex4}

\usepackage[usenames]{color}
\usepackage{amssymb,epsfig}
\usepackage{psfrag}

\renewcommand{\theequation}{\arabic{section}.\arabic{equation}}

\newcommand{\beq}[1]{
\begin{equation}\label{#1}}
\newcommand{\eeq}{\end{equation}}
\newcommand{\bea}[1]{
\begin{eqnarray}\label{#1}}
\newcommand{\eea}{\end{eqnarray}}

\newcommand{\insertfig}[2]{\mbox{\epsfxsize=#1cm \epsfbox{#2.eps}}}

\begin{document}



\preprint@{
\hspace{2.2mm}IPPP/07/54\\
\phantom{abc}DCPT/07/108\\
}
\vspace{4mm}

\title{Exclusive processes in position space
            and the pion distribution amplitude}

\author{V. M. Braun\\}

\affiliation{Institut f{\"u}r Theoretische Physik, Universit{\"a}t
          Regensburg, D-93040 Regensburg, Germany\\}

\author{D. M{\"u}ller\\}

\affiliation{Institut f{\"u}r Theoretische Physik II,
Ruhr-Universit{\"a}t Bochum, D-44780 Bochum, Germany\\}

\begin{abstract}
\noindent We suggest to carry out  lattice calculations of current
correlators in position space, sandwiched between the vacuum and a
hadron state (e.g.~pion), in order to access hadronic light-cone
distribution amplitudes (DAs). In this way the renormalization
problem for composite lattice operators is avoided altogether, and
the connection to the DA is done using perturbation theory in the
continuum. As an example, the correlation function of two
electromagnetic currents is calculated to the
next-to-next-to-leading order accuracy in perturbation theory and
including the twist-4 corrections. We argue that this strategy is
fully competitive with direct lattice measurements of the moments
of the DA, defined as matrix elements of local operators, and
offers new insight in the space-time picture of hard exclusive
reactions.
\end{abstract}


\maketitle

\section{Introduction}

Hadron light-cone distribution amplitudes (DAs) present the
principal nonperturbative input in the pQCD description of hard
exclusive reactions and are to a large extent complementary to the
usual parton distributions. The existing information on DAs is,
however, very limited. The main reason for this is that the DAs
are much more difficult to access experimentally: for realistic
momentum transfers, the contributions of interest are often
contaminated by large nonperturbative corrections coming from the
end-point regions in the quark momentum fraction.

The pion leading twist DA is the simplest one and has attracted
most attention. There is increasing evidence that at the scale of
the order of 1 GeV this DA differs considerably from its
asymptotic form. In particular, QCD sum rule estimates
\cite{Bakulev:2001pa,Bakulev:2002hk}, lattice calculations
\cite{Braun:2006dg} and the analysis of the experimental data on
the transition form factor $\gamma^*\to\pi\gamma$
\cite{SchYak99,Bakulev:2002uc,Bakulev:2001pa} are consistent with
the positive value of the second Gegenbauer moment of the pion DA
which is roughly a factor two below the original estimate by
Chernyak and Zhitnitsky \cite{Chernyak:1981zz}. Beyond the second
moment, very little is known. The analysis of the $ \gamma^* \to
\pi \gamma $ form factor in Ref.~\cite{SchYak99,Bakulev:2001pa}
indicates a negative value for the fourth Gegenbauer moment, but
the status of this result is not clear as the analysis has some
model dependence. The lattice calculations of the fourth and
higher Gegenbauer moments would be very difficult because they
contribute with a small coefficient, and because the lattice
renormalization of local operators with many derivatives becomes
too cumbersome. The aim of this letter is to suggest an
alternative approach, based on the lattice calculation of
exclusive amplitudes in coordinate space. We will argue that the
interpretation of such calculations in the framework of QCD
factorization is equally straightforward and offers new insights
compared to the standard momentum space formulation. {}From the
lattice side, the advantage is that the renormalization problem
for composite operators is avoided altogether, but, instead, in
order to be sensitive to the detailed structure of the pion DA,
one needs pion sources with large momentum,  at least of order
2--3 GeV.

The idea to  emphasis  the coordinate rather than momentum-space
dependence of the correlation functions is by itself not new, see
e.g.~\cite{Schafer:1995uz,Braun:1994jq}.  Our proposal goes in the
same direction as the work \cite{Aglietti:1998ur,Abada:2001yt},
the difference is that we suggest to calculate physical
observables that are free from renormalzation uncertainties. Also,
we demonstrate that the analysis of such correlation functions in
continuum theory is aided by the conformal operator product
expansion.

The presentation is organized in follows. In Sec.~\ref{Sec-Exa} we
introduce the coordinate-space correlation function of two
electromagnetic currents sandwiched between vacuum and the pion.
We calculate this correlation function to leading order (LO) in
QCD perturbation theory and discuss the physical interpretation.
In Sec.~\ref{Sec-StaArt} the state-of-the-art
(next-to-next-to-leading order) calculation of this correlation
function is presented, including two-loop radiative corrections
and nonperturbative twist-4 effects. The main result of this
analysis is that the QCD corrections remain well under control for
all pion momenta. In Sec.~\ref{Sec-RevDA} we discuss possible
strategies and the potential accuracy of the extraction of the
pion distribution amplitude from the coordinate space dependence
of this correlation function, assuming that it is measured in a
limited range of distances accessible in lattice calculations. The
corresponding setup and several possible generalizations are
discussed in Sec.~\ref{Sec-Lat}, which also contains a summary and
our final conclusions.

\section{Example}
\label{Sec-Exa} \setcounter{equation}{0}
As an example, we consider the correlation function of two
electromagnetic currents sandwiched between vacuum and the pion
state
\begin{equation}
T_{\mu\nu} = \langle 0| T\{j_\mu(x)j_{\nu}(-x)\}\pi^0(p)\rangle\,,
\label{lo:1}
\end{equation}
where
\begin{eqnarray}
j_\mu(x) =
\frac{2}{3}\bar u(x) \gamma_\mu u(x) -
\frac{1}{3} \bar d(x) \gamma_\mu d(x)\,,
\end{eqnarray}
which is the coordinate space analog of the pion transition form
factor involving two photons. Note that one and the same
correlation function (\ref{lo:1}) enters the pion decay $\pi^0\to
\gamma\gamma$ and the form factors $\gamma^*\to\pi^0\gamma^*$ for
two virtual or $\gamma^*\to\pi^0\gamma $ for one virtual and one
real photons. Differences arise at the stage when one goes over to
the states with given momenta: depending on the virtuality,
different coordinate space regions are emphasized/suppressed by
the Fourier transform.

At small space-like separations $|x^2| \ll
1/\Lambda^2_{\mathrm{QCD}}$ the amplitude in (\ref{lo:1}) can be
calculated using the operator product expansion (OPE). To
LO in the strong coupling,
\begin{figure}[ht]
\begin{center}
\mbox{
\begin{picture}(250,105)(0,0)
\put(35,-5){\insertfig{6}{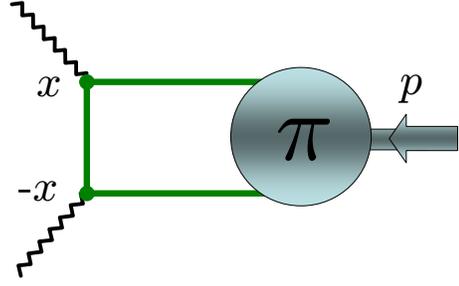}}
\end{picture}
}
\end{center}
\caption{\label{figsum} \small Leading-order contribution to the
correlation function (\ref{lo:1}) }
\end{figure}
the answer is obtained from the diagram in Fig.~1 and reads:
\begin{equation}
T_{\mu\nu} =
-\frac{5i}{9}f_\pi\epsilon_{\mu\nu\rho\sigma}
\frac{x^\rho p^\sigma}{8\pi^2 x^4}T(p\cdot x,x^2)
\label{def:T}
\end{equation}
with
\begin{equation}
T(p\cdot x,x^2) = \int_0^1 du\, e^{i(2u-1)p\cdot x} \phi_\pi(u,\mu)
\label{def:T-0}\,.
\end{equation}
The relevant nonperturbative input is provided by the pion DA
defined as
\begin{eqnarray}
\lefteqn{
\langle 0| \bar q(a_1 n) \gamma_\mu\gamma_5 q(a_2 n)\pi^0(p)\rangle=}
\nonumber
\\&&{} =
i f_\pi p_\mu \int_0^1 du e^{-ipn(\bar u a_1 + u a_2)}\phi_\pi(u,\mu)\,,
\end{eqnarray}
where $f_\pi =93$~MeV is the pion decay constant, $\bar u\equiv
1-u$, $n_\mu$ is a light-light vector, $n^2=0$, and $\mu$ is the
normalization scale, which is set to be of the order
$1/\sqrt{|x^2|}$.

In the following we write the pion DA as an expansion over
Gegenbauer polynomials,
\begin{equation}
 \phi_\pi(u,\mu) =
 6u\bar u \sum_{n=0}^\infty \phi_n(\mu) C^{3/2}_n(2u-1)\,,
\label{GegExp}
\end{equation}
and use the fact that the Fourier transform
\begin{eqnarray}
\label{Def-FT-Geg} \lefteqn{
 8\int_{0}^1 du\, u\bar u\, e^{i\rho (2u-1)} C^{3/2}_n(2u-1) =}
\nonumber\\&&{}=
   \sqrt{2\pi} (n+1)(n+2)i^n \rho^{-3/2} J_{n+3/2}(\rho)
\end{eqnarray}
leads to Bessel functions. To our accuracy we obtain then for the
correlation function (\ref{def:T-0})
\begin{equation}
 T(p\cdot x,x^2) =
 \frac34\sum_{n=0}^\infty \phi_n(\mu) {\mathcal F}_n\left(p\cdot x\right)
 \label{def:T-1}
\end{equation}
with
\begin{equation}
\label{Def-F-LO}
 {\mathcal F}_n(\rho) = i^n \sqrt{2\pi}\frac{(n+1)(n+2)}{2} \rho^{-3/2} J_{n+3/2}(\rho)\,.
\end{equation}
We will view this result (\ref{def:T-1}) as a partial wave
expansion of the correlation function $T(\rho=p\cdot x,x^2)$. (a
group theoretical explanation will be given below). The partial
waves ${\mathcal F}_n(\rho)$, expressed by Bessel functions
$J_{n+3/2}(\rho)$ with half integer index, are simply given in
terms of trigonometric functions, e.g., we have for $n=0$:
${\mathcal F}_0(\rho)=2[\sin(\rho)-\rho \cos(\rho))/\rho^3$. In
particular this term, appearing as the leading one in the
expansion (\ref{def:T-1}), corresponds to the contribution of the
asymptotic distribution amplitude, while ${\mathcal F}_2(\rho)$
and ${\mathcal F}_4(\rho)$ take into account contributions of the
second, $C^{3/2}_2(2u-1)$, and the fourth, $C^{3/2}_4(2u-1)$,
Gegenbauer moments, respectively. Note that only even values of
$n$ contribute to the series because of $C$-parity conservation.
The used approximation allows for a partonic interpretation of the
correlation function $T(p\cdot x,x^2)$, which is obvious: it
corresponds to a probability amplitude of the valence quark
distribution in the pion in the longitudinal distance
(``Ioffe-time'') $\rho=p\cdot x$ at transverse distance $x^2$.

\begin{figure}[ht]
\begin{center}
\mbox{
\begin{picture}(250,105)(0,0)
\put(0,35){\rotatebox{90}{$ i^n {\cal F}_n(\rho)$}}
\put(15,-10){\insertfig{6.4}{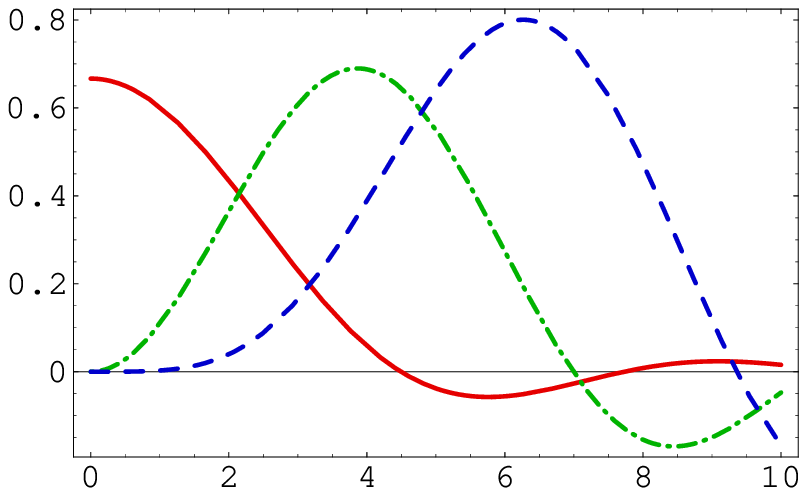}} \put(190,-15){$\rho$}
\end{picture}
}
\end{center}
\caption{\small From left to right: ${\mathcal F}_0(\rho)$
(solid), $-{\mathcal F}_2(\rho)$ (dashdotted), ${\mathcal
F}_4(\rho)$ (dashed).}
 \label{figharm}
\end{figure}

The partial waves ${\mathcal F}_0(\rho)$, $-{\mathcal F}_2(\rho)$
and ${\mathcal F}_4(\rho)$ are plotted in Fig.~\ref{figharm}.
These oscillating functions are strongly peaked at a position that
moves with increasing index $n$ to the r.h.s.~and so the
contribution of the asymptotic DA and the corrections are well
separated. On the other hand, large values of $p\cdot x$ and,
hence, of the pion momentum are needed to probe higher-order
Gegenbauer moments efficiently.

\begin{figure}[ht]
\begin{center}
\mbox{
\begin{picture}(250,105)(0,0)
\put(0,40){\rotatebox{90}{$T(\rho,x^2)$}}
\put(15,-10){\insertfig{6.4}{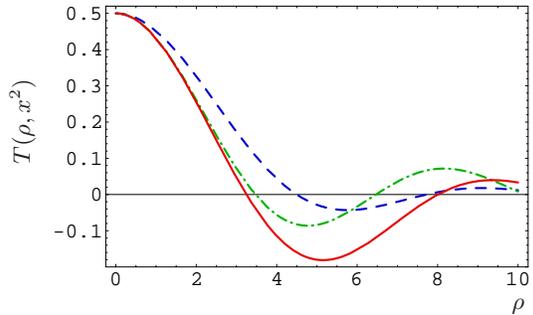}} \put(190,-15){$\rho$}
\end{picture}
}
\end{center}
\caption{ \label{Fig-LO} \small The correlation function
$T(\rho,x^2)$, leading-order (\ref{def:T-1}), calculated using
asymptotic pion DA (dashed), the model with $\phi_2=0.25$ and two
choices $\phi_4 = 0.1$ (dashdotted) and $\phi_4 = -0.1$ (solid). }
\end{figure}

\begin{figure}[ht]
\begin{center}
\mbox{
\begin{picture}(250,105)(0,0)
\put(0,40){\rotatebox{90}{$\phi(u)$}}
\put(15,-10){\insertfig{6.4}{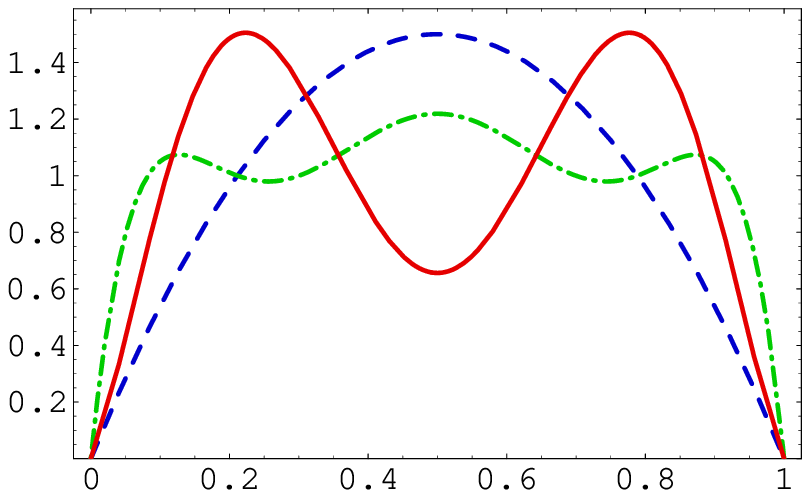}} \put(180,-15){$u$}
\end{picture}
}
\end{center}
\caption{ \label{Fig-LO-DA} \small Models for the pion DA as
specified in Fig.~\ref{Fig-LO}.}
\end{figure}

{}To see this, we plot in Fig.~\ref{Fig-LO} the correlation
function $T(\rho,x^2)$, given in the approximation
(\ref{def:T-1}), as a function of $\rho=p\cdot x$ for the
asymptotic pion DA (dashed curve), and for the model with
$\phi_2=0.25$ and two choices $\phi_4 = 0.1$ and $\phi_4 = -0.1$,
respectively, cf.~Fig.~\ref{Fig-LO-DA}.

Note that the value of $T(\rho,x^2)$ at $\rho=0$ is equal to 1/2
in our normalization (up to radiative and higher twist corrections
discussed in the next Section) and does not depend on the shape of
the pion DA. The deviation from the asymptotic DA (dashed curve)
for a realistic model is significant starting $\rho \sim 1$.  The
dependence on $\phi_4$, taken as $\phi_4 = 0.1$ (dashdotted curve)
and $\phi_4 = -0.1$ (solid curve), becomes visible at distances
$\rho \ge 3-4$. We remark that a more closer look, given in
Sect.~\ref{Sec-RevDA}, reveals that a partial wave analysis might
allow to access the forth Gegenbauer moment even for $\rho
\lesssim 3$. With increasing $\rho$ the correlation function
changes sign and becomes negative, unless the fourth Gegenbauer
moment is larger as the second one, which seems to be unlikely
(not shown).

The $\rho$-behavior of the models, plotted in Fig.~\ref{Fig-LO},
results from the $\cos$-Fourier transform (\ref{def:T-0}) of the DA
(because for a symmetric DA the $\sin$-component dies out) and can
qualitatively be understood as follows. The position of the first
zero at $\rho= \rho_{0}$ gives us the effective width
$\pi/\rho_{0}$ of the DA in momentum space. For instance, a flat
DA $\phi_\pi(u)=1$ corresponds to $\rho_{0}=\pi$, an equal
momentum sharing DA $\phi_\pi(u)=\delta(u-1/2)$ leads to a
constant correlation function, i.e., $\rho_0\to \infty$, while the
effective width of the asymptotic DA follows from $\rho_{0}
\approx 1.43 \pi.$ Compared to the asymptotic DA, for the
realistic models, shown by the solid and the dash-dotted curves,
the first zero is shifted to the left, which indicates that these
DAs are broader than the asymptotic one, cf.~Fig.~\ref{Fig-LO-DA}.
The difference between the solid and the dash-dotted models, on
the other hand, is mostly pronounced near the first minimum. The
magnitude of the correlation function at this minimum
$\rho=\rho_1$ is a measure for the overlap of the DA with the
harmonic $\cos((2u-1)\rho_{1})$ which is maximized at $u_{1} =
1/2-\pi/\rho_{1}\sim 0.2$. It is large for the models with a
two-hump DA like the one of Ref.~\cite{Bakulev:2001pa}, where it
was also argued that all higher Gegenbauer moments are negligible.
It is much smaller for a dash-dotted model corresponding to a flat
(albeit still oscillating) DA as shown in Fig.~\ref{Fig-LO-DA}.
These oscillations are not significant and can be understood as an
artifact of truncating the conformal spin partial wave expansion.
In fact, a conjectured ADS/QCD model $\phi^{\rm ADS/QCD}= (8/\pi)
\sqrt{u(1-u)}$ \cite{Polchinski:2002jw,Brodsky:2003px} gives rise
to a similar shape.
We add to our discussion that very narrow DA's
and those with nodes, leading to huge resonance effects,  are
clearly distinguishable in the correlation function from those
shown in Fig.~\ref{Fig-LO}, even at smaller values of $\rho$.

\section{State-of-the-Art }
\label{Sec-StaArt}

The previous discussion was mainly qualitative. In this section we
present the state-of-the-art calculation of the correlation
function $T(p\cdot x,x^2)$, defined in Eq.~(\ref{def:T}), in
next-to-next-to-leading (NNLO) of perturbative QCD and including
the twist-four corrections.

An adequate technical framework is provided by the conformal
operator product expansion (OPE). Recall that the expansion in
Gegenbauer polynomials in (\ref{GegExp}) alias the partial wave
decomposition in (\ref{def:T-1}) to LO are governed by conformal
symmetry: the expansion is organized as the irreducible
decomposition of the product of two currents into conformal
operators. The partial waves specified in Eq.~(\ref{Def-F-LO}) are
nothing but the LO Wilson-coefficients of the leading
twist operators in the conformal OPE.  They can be viewed as the
Clebsch--Gordon coefficients of the collinear conformal group
$SL(2,\mathbb{R})$  and are labelled by the conformal spin
$j=n+2$, for a review see Ref.~\cite{BraKorMue03}. The advantage
of such a decomposition is that the Gegenbauer moments do not mix
under evolution in LO or, in other words, the conformal spin is a
good quantum number to this accuracy.

Taking into account radiative corrections to the leading twist
Wilson-coefficients translates to the modification of the partial
waves which become scale- (and scheme-) dependent:
$$ {\cal F}_n(\rho)\to {\cal F}_n(\rho,-\mu^2 x^2;\alpha_s(\mu))\,.$$
In turn, the higher-twist corrections  yield $x^2$ suppressed
additive terms and for our present purposes it is convenient to
reexpand them in terms of the leading-twist partial waves taken to
leading-order. We end up with the expansion of the form
\begin{eqnarray}
\label{Def-T}
&& T(\rho,x^2)=
\\
&&=\frac{3}{4}\sum_{n=0}^\infty
\phi_n(\mu^2)\, {\cal F}_n(\rho,-\mu^2 x^2;\alpha_s(\mu)) +
O(\alpha_s^3)
\nonumber\\
&&\;\; +    \frac{3}{4} x^2 \sum_{n=0}^\infty \phi^{(4)}_n(\mu^2)\,
{\cal F}_n(\rho) + O(x^2 \alpha_s)+ O(x^4)\,,
\nonumber
\end{eqnarray}
where $\phi^{(4)}_n$ are related to the matrix elements of
twist-four operators, weighted with specific Wilson coefficients.
Viewed this way, both the corrections to the partial waves and the
twist-4 coefficients $\phi^{(4)}_n(\mu^2)$ are specific for the
considered correlation function. The Gegenbauer moments
$\phi_n(\mu^2)$ are universal, however, they depend on the scheme
conventions.

In what follows we consider separately perturbative and twist-four
corrections in some detail.

\subsection{Perturbative corrections}
\label{Sec-PerCor}

The next-to-leading (NLO) perturbative corrections to the pion
transition form factor were calculated in the minimal subtraction
($\overline{\rm MS}$) scheme in Refs.~\cite{AguCha81,Bra83,KadMikRad86}
and were supplemented by the evaluation of the logarithmical scale
change \cite{DitRad84,Sar84,MikRad85,Mue94}.
In addition NNLO diagrams that are proportional to $n_f$, the number
of light quark flavors, have been evaluated in the same scheme \cite{MelNicPas02}. This
result can be used to obtain the NNLO corrections that are
proportional to $\beta_0 = (11/3) C_A - (2/3) n_f$, the first
coefficient appearing in the QCD beta-function $\beta(g)/g =
-(\alpha_s/4\pi^2)\beta_0 + O(\alpha_s^2)$. Finally, the
constraints imposed by the conformal symmetry, tested at NLO level
\cite{Mue97a,BelMue97a}, have been used to obtain the missing
terms at NNLO. A detailed NNLO analysis can be found in
Ref.~\cite{MelMuePas02}, see also Appendix \ref{AppPerQua} for a
discussion of different factorization schemes. A Fourier transform
of these results provides one with the radiative corrections to
the correlation function (\ref{def:T}) in position space.

It is instructive to consider the results in the hypothetical
conformal limit, in which the $\beta$ function is vanishing (which
means, technically, that the $\beta_0$ proportional terms are
omitted). In this case the modification of the partial waves, cf.
Eq.~(\ref{Def-T}), is entirely governed by conformal symmetry:
\begin{eqnarray}
\label{Def-ParWav-CS}
 {\cal F}_n &=&
C_{n}(\alpha_s)\; (-\mu^2 x^2)^{\frac{\gamma_n}{2}}  i^n
\sqrt{\pi}  \frac{(n+1) (n+2)}{4}
\\
&&\!\!\!\times \frac{\Gamma(n+5/2+\gamma_n/2)}{\Gamma(n+5/2)}
\left(\frac{\rho}{2}\right)^{-\frac{3+\gamma_n}{2}}
J_{n+\frac{3+\gamma_n}{2}}(\rho).
\nonumber
\end{eqnarray}
As compared to the LO expressions  (\ref{Def-F-LO}), the  $\rho$
dependence  is modified by the anomalous dimensions
$\gamma_n(\alpha_s)$, which alter, e.g., the index of Bessel
functions. In addition the normalization is changed by the factor
\begin{equation}
\label{Def-ParWav-CS-coe} C_{n}(\alpha_s) =
\frac{\Gamma(2-\gamma_n/2) \Gamma(1+n)}{\Gamma(1+n+\gamma_n/2)}
c_{n}(\alpha_s)\,,
\end{equation}
where $c_{n}(\alpha_s)$ are the Wilson coefficients that appear in
polarized deep inelastic scattering structure function $g_1$ and
are known to NNLO \cite{ZijNee92}. Since of conformal symmetry, a
scale change does not lead to a mixing of conformal partial waves
(\ref{Def-ParWav-CS-coe}) or the Gegenbauer moments.

One possibility to go beyond the conformal limit is to restore
both the scale dependence of the coupling and the renormalization
logs within the normalization condition of
Ref.~\cite{MelMuePas02}. This scheme is discussed in Appendix
\ref{AppPerQua}; we refer to it as the conformal subtraction
scheme, the $\overline{\rm CS}$ scheme. We will use the
renormalization group improved partial waves, defined in
Eq.~(\ref{Def-PW-CS-NNLO}), for the numerical analysis. In
general, the anomalous dimensions govern the evolution of the
Gegenbauer moments with respect to a scale change:
\begin{equation}
\mu \frac{d}{d\mu} \phi_n(\mu^2) = -\gamma_n(\alpha_s(\mu))
\phi_n(\mu^2) + \cdots .
\end{equation}
Even in the $\overline{\rm CS}$ scheme which is designed to make
maximal use of the conformal symmetry, we expect that the
Gegenbauer moments will mix under evolution, indicated by the
ellipsis.  This mixing is induced by the trace anomaly and gives
rise to a $(\alpha_s/2\pi)^2 \beta_0 \ln(-x^2 \mu^2_0) $
proportional contribution which, however, vanishes at a reference
scale $-x^2=1/\mu_0^2$. We expect that this unknown NNLO mixing
effect is small and can safely be ignored.

The partial waves in the conventional $\overline{\rm MS}$ scheme
can be obtained from those in  $\overline{\rm CS}$ by the
appropriate transformation. In particular the Gegenbauer moments
in these two schemes are related as
\begin{equation}
\label{rotMSCS} \phi^{\overline{\rm CS}}_n = \phi^{\overline{\rm
MS}}_n - \frac{\alpha_s}{2 \pi} \sum_{m=0}^{n-2}  B^{(1)}_{n m}
\phi^{\overline{\rm MS}}_m + O(\alpha_s^2)\,.
\end{equation}
The matrix $B^{(1)}_{nm}$ is known explicitly and is given in
Eq.~(\ref{Def-B}) of Appendix \ref{AppPerQua}. This relation is
also valid in the hypothetical conformal limit.  Note that the
$\overline{\rm MS}$ Gegenbauer moments are given  by a finite sum
of those in the $\overline{\rm CS}$. Consequently, plugging
Eq.~(\ref{rotMSCS}) into Eq.~(\ref{Def-T}), one sees that in the
$\overline{\rm MS}$ scheme all conformal partial waves
(\ref{Def-ParWav-CS}) are excited even if a truncated model for
the DA is used. We stress that the scheme independence of the
correlation function is only guaranteed  if the Gegenbauer moments
are rotated at the given input scale. Taking the same model in
different schemes will in general lead to different predictions
for the correlation function.

As we will see in Sect.~\ref{Sec-RevDA}, such scheme--dependent
mixing effects are numerically important. For illustration,
consider the asymptotic DA which we define here as the zero mode
of the evolution kernel in a given scheme and order of
perturbation theory. By construction, the asymptotic DA does not
evolve under scale transformations. In the $\overline{\rm CS}$
scheme to NLO this amounts to the same choice of the Gegenbauer
coefficients as in LO: $\phi_0=1$ and $\phi_n=0$ for $n>0$. The
one-loop corrections to the correlation function (\ref{Def-T}) are
then given by
\begin{equation}
\label{Def-T-CS-asy} T^{\rm as}\simeq \frac{3}{4} \left[1-
\frac{\alpha_s(\mu)}{\pi} + {\cal O}(\alpha_s^2)\right]\,
{\cal F}_0(\rho)\,.
\end{equation}
The current conservation implies $\gamma_0(\alpha_s)=0$,
protecting the $\rho$-dependence of the lowest partial wave from
radiative corrections. In the $\overline{\rm MS}$ scheme, on the
other hand, the Gegenbauer moments of the asymptotic DA for
$n\ge2$ are nonzero and are obtained by the rotation
(\ref{rotMSCS}):
\begin{eqnarray}
\label{AsyDA-MS} \phi^{{\rm as}-\overline{\rm MS}}_n &=&
\frac{\alpha_s(\mu)}{2 \pi} \frac{8 C_F (2 n+3)}{n (n+1) (n+2) (n+3)}
\\
&&\times \left(S_1(n+2)-\frac{n+3}{2(n+1)}\right)
+{\cal O}(\alpha_s^2)\,, \nonumber
\end{eqnarray}
whereas the normalization, i.e., $\phi_0=1$, remains unchanged.
The excitation of higher Gegenbauer moments is, however, rapidly
decreasing with $n$: For  $n=\{2,4,6,8\}$ we find the values
$\{0.124 , 0.039 , 0.018 , 0.009 \}$ in units of $\alpha_s(\mu)$,
respectively. A more extensive discussion about scheme
dependence can be found in Ref.\ \cite{Mue9498}.

Finally, the limit $\rho\to 0$ of the correlation function
(\ref{def:T}) yields the analog of the Bjorken sum rule in
polarized deep inelastic scattering:
\begin{equation}
\label{SumRul} T(\rho= 0, x^2)\simeq \frac{1}{2} C_{0}(-\mu^2 x^2,
\alpha_s(\mu))\,.
\end{equation}
In this limit, which is independent on the factorization scheme,
any information contained in the DA is washed out. Hence, this sum
rule is a {\em pure} QCD prediction, which is known as
perturbative expansion to three loop order \cite{LarVer91}.

After these general remarks, we consider the perturbative
corrections for the particular models of the pion DA specified in
Sect.~\ref{Sec-Exa}. We assume that the models are defined at the
scale $1\,$GeV and set the factorization scale to $\mu^2 =
-1/x^2$. For the demonstration we consider two current
separations: a larger one $-x^2=1/{\rm GeV}^{2}\;\left[\approx
(0.2\, {\rm fm})^2\right]$ and a smaller one $-x^2=1/4{\rm
GeV}^{2}\; \left[\approx (0.1\, {\rm fm})^2\right]$. In the latter
case evolution of Gegenbauer moments is taken into account: The
leading logs are resummed and non--leading ones are consistently
combined with the $\alpha_s/2\pi$ power expansion of the
correlation function (\ref{Def-T}), for details see, e.g.,
Ref.~\cite{MelMuePas02}. The coupling is specified at the
normalization point $\mu=2\, {\rm GeV}$ to be $\alpha_s=0.36$ and
$\alpha_s=0.3$ at LO and NLO, respectively.

\begin{figure}[ht]
\begin{center}
\mbox{
\begin{picture}(250,220)(0,0)
\put(0,128){\rotatebox{90}{$T(\rho,x^2 = -1/4{\rm GeV}^{2} )$}}
\put(15,98){\insertfig{7}{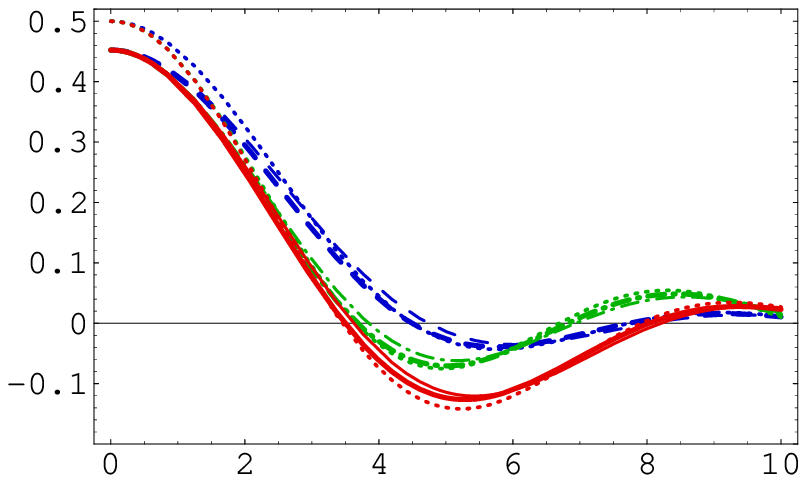}}
\put(0,20){\rotatebox{90}{$T(\rho,x^2 = -1/{\rm GeV}^{2} )$}}
\put(15,-10){\insertfig{7}{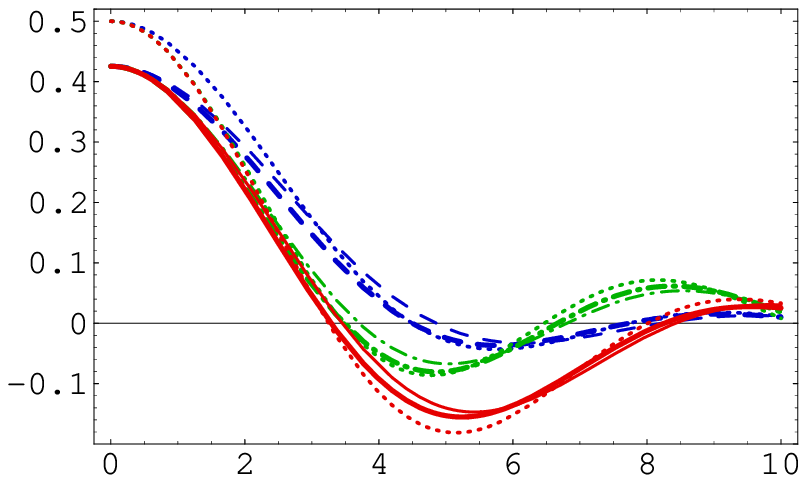}} \put(180,-15){$\rho=p\cdot
x$}
\end{picture}
}
\end{center}
\caption{ \label{Fig-MScorPT} \small NLO predictions for the
correlation function (\ref{def:T}) in the $\overline{\rm MS}$
scheme (thin) and rotated to the $\overline{\rm CS}$ one (thick)
at the scale $-x^2= 1/{\rm GeV}^{2}$  (lower panel) $-x^2=
1/4\,{\rm GeV}^{2}$ (upper panel) compared to the LO results
(dots) for the three models of the pion DA shown in
Fig.~\ref{Fig-LO-DA}.}
\end{figure}

The NLO results are shown in Fig.~\ref{Fig-MScorPT} for $-x^2 =
1/{\rm GeV}^2$ and $-x^2 =1/4{\rm GeV}^2$ in the lower and
upper panel, respectively. Notice that the perturbative
corrections lead to a universal shift of the value at $\rho=0$,
which is 1/2 in LO, downwards to $1/2(1-\alpha_s/\pi)$ which is
about -15\% for the larger and -9\% for the smaller  separation
between the currents, respectively. This effect is entirely due to
the negative NLO correction in the sum rule (\ref{SumRul}).

Besides the overall normalization, perturbative corrections result
in a slight change of the $\rho$ dependence, seen as a shift of
zeros and extrema to larger $\rho$ values (compare thin and dotted
lines). This scheme dependent effect arises from the excitation of
higher conformal partial waves, as discussed above, and is more
pronounced at larger separations  (lower panel). Note that the
$\rho$-dependence corresponding to the asymptotic DA (dashed
curves), evaluated in the $\overline{\rm CS}$ scheme (thick
curves), is not affected, see Eq.~(\ref{Def-T-CS-asy}). For the
other two models in the $\overline{\rm CS}$ scheme only a slight
shift of extrema appears, which is visible in the lower panel at
larger values of $\rho$ (compare thick and dotted lines). A more
quantitative look reveals that in general the absolute size of
radiative correction grows somewhat with increasing $\rho$.

Note that at smaller distances the model dependence slightly
weakens, compare the magnitude of solid or dash-dotted curves at
the first minima in the upper and lower panel. This is caused by
evolution, which does not change the value at $\rho=0$, however,
reduces Gegenbauer moments with a strength that grows with
increasing conformal spin. Obviously, even at $-x^2 = 1/4{\rm
GeV}^2$ the model predictions remain clearly distinguishable.

The NNLO corrections evaluated in the $\overline{\rm CS}$ scheme
(not shown) lead to a further decrease of the normalization, so
that the net reduction for $\rho=0$ is, compared to LO,  $-25\%$
and $-13\%$ for $-x^2 = 1/{\rm GeV}^2$ and $-x^2 =1/4{\rm GeV}^2$,
respectively. The modification  of the $\rho$ dependence is,
however, negligible as compared to the NLO result in the
$\overline{\rm CS}$ scheme, and is hardly visible even for the
larger separation $-x^2 = 1/{\rm GeV}^2$. As mentioned above,
mixing effects due to the NNLO evolution are not taken into
account, but are expected to be tiny.

\subsection{Higher-twist Corrections}
Higher--twist corrections are in general suppressed by full powers
of the separation $x^2$ between the currents. In particular,
including twist--four contributions one obtains
\begin{equation}
\label{DefT-Tu}
 T(p\cdot x,x^2) = \int_0^1\! du\, e^{i(2u-1) p\cdot x} t(u,x^2)\,,
\end{equation}
where
\begin{eqnarray}
\label{DefT-Tw4}
 t(u,x^2) &=& \phi_\pi(u) +
 \frac{x^2}{4}\phi_4(u) -
 x^2
 \int_0^u\! d\alpha_1\!\!\int_0^{\bar u}\! d\alpha_2
\nonumber\\
&& \times \left[
\frac{1}{\alpha_3}\widetilde\Phi_4(\underline{\alpha}) +
 \frac{(u-\bar u-\alpha_1+\alpha_2)}{\alpha_3^2}
 \Phi_4(\underline{\alpha})\right]\,.
 \nonumber\\
 {\ }
\end{eqnarray}
The leading--twist perturbative corrections are not shown for
brevity. The notations correspond to Ref.~\cite{Ball:2006wn},
where one can find the definitions and the corresponding
expressions for the twist-four distribution amplitudes to the
next-to-leading conformal spin accuracy. In the last term
$\alpha_1, \alpha_2$ and $\alpha_3=1-\alpha_1-\alpha_2$ are the
quark, antiquark and gluon momentum fractions, respectively.
Including the contributions of the lowest and the next-to-lowest
conformal spin one obtains \cite{BraFil90,Ball:2006wn}
\begin{eqnarray}
\label{DefDA-Tw4}
  \phi_4(u) &=& \frac{200}{3}\delta^2_\pi u^2 \bar u^2 +
  21 \delta^2_\pi \omega_{4\pi} \Big\{ u\bar u (2+13 u\bar u)
\nonumber\\
  &&+ [2u^3(6u^2-15u+10)\ln u] + [u\leftrightarrow \bar u] \Big\}\,,
\nonumber\\
   \widetilde\Phi_4(\underline{\alpha}) &=&
   120 \alpha_1\alpha_2\alpha_3 \delta^2_\pi
 \left[-\frac{1}{3}+\frac{21}{8}\omega_{4\pi} (3\alpha_3-1)\right]\,,
\nonumber\\
 \Phi_4(\underline{\alpha}) &=&
 120\alpha_1\alpha_2\alpha_3 (\alpha_1-\alpha_2)
 \delta^2_\pi \frac{21}{8} \omega_{4\pi}\,.
\end{eqnarray}
The nonperturbative parameters $\delta^2_\pi$ and $\omega_{4\pi}$
are defined as reduced matrix elements of local operators, for
example
\begin{equation}
 \langle 0|\bar q ig\widetilde G_{\mu\nu}\gamma_\nu q |\pi^0(p)\rangle
 = - f_\pi \delta^2_\pi p_\mu\,.
\end{equation}
Numerical estimates for these matrix elements are available from
QCD sum rules:
\begin{equation}
\label{Tw4-par}
  \delta^2_\pi = (0.18\pm 0.06)~\mbox{GeV}^2\,,\qquad
  \omega_{4\pi} = 0.2\pm 0.1
\end{equation}
at the scale 1 GeV [$\delta^2_\pi = (0.14\pm 0.05)~\mbox{GeV}^2$,
$\omega_{4\pi} = 0.13\pm 0.07$ at the scale 2 GeV], see
Ref.~\cite{Ball:2006wn} and the references therein.

Note that beyond the leading conformal spin accuracy the
twist-four contributions are not polynomials \cite{BraFil90}.
Plugging Eq.~(\ref{DefDA-Tw4}) into Eq.~(\ref{DefT-Tw4}), we find,
however, that the next-to-leading conformal spin contributions
completely cancel each other and the leading conformal spin yields
the total $O(x^2)$ correction:
\begin{equation}
t(u,x^2) =  \phi_2(u)+x^2 \frac{8}{9} \delta_\pi^2 \cdot 30
u^2\bar u^2\,.
\end{equation}
This can be effectively rewritten as the correction to the first
two leading twist Gegenbauer coefficients $1\to 1+(8/9)
x^2\delta_\pi^2$, $\phi_2\to \phi_2 - (8/9)(1/6) x^2\delta_\pi^2$.
Hence, we have
\begin{equation}
\phi^{(4)}_0(\mu) = \frac{8}{9} \delta_\pi^2(\mu)\,,\quad
\phi^{(4)}_2(\mu) = -\frac{8}{54} \delta_\pi^2(\mu)\,.
\label{effect4}
\end{equation}
The upper bound for the contributions to (\ref{DefDA-Tw4}) of
higher conformal spins can be obtained using the renormalon model
of Ref.~\cite{Braun:2004bu}. These extra contributions mainly
influence the large $\rho$-behavior of the correlation function
while the  numbers in (\ref{effect4}) are not affected. We see
that, to LO accuracy, higher--twist correction produce an additive
shift in the physical values of the Gegenbauer moments of the pion
DA, which is calculable, at least in principle.

\begin{figure}[ht]
\begin{center}
\mbox{
\begin{picture}(250,220)(0,0)
\put(0,128){\rotatebox{90}{$T(\rho,x^2 = -1/4{\rm GeV}^{2} )$}}
\put(15,98){\insertfig{7}{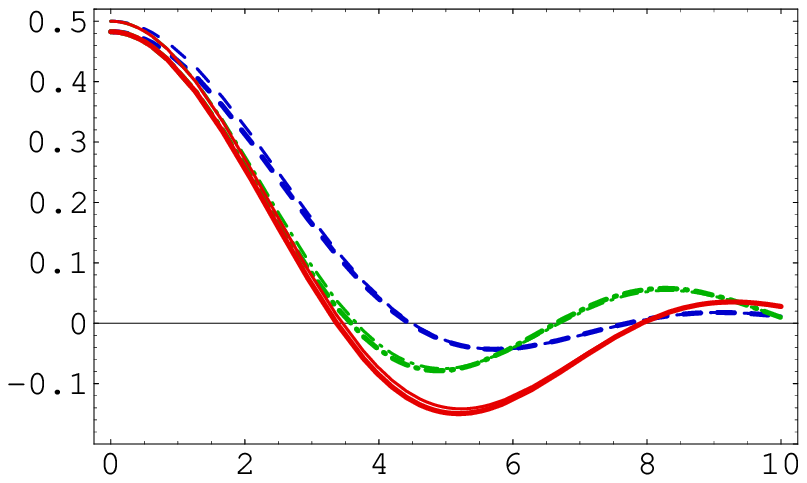}}
\put(0,20){\rotatebox{90}{$T(\rho,x^2 = -1/{\rm GeV}^{2} )$}}
\put(15,-10){\insertfig{7}{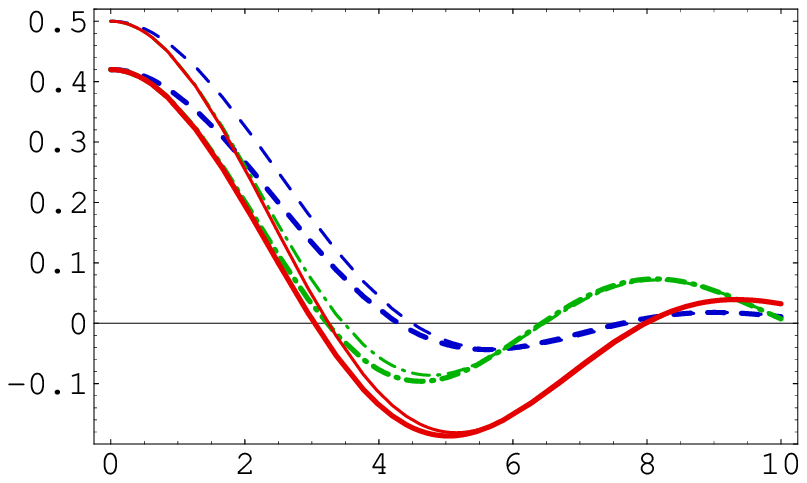}} \put(180,-15){$\rho=p\cdot
x$}
\end{picture}
}
\end{center}
\caption{ \label{Fig-corTW4} \small LO predictions for the
correlation function (\ref{def:T}) with (thick) and without (thin
curves) twist-four corrections for two choices of the
quark-interquark separation: $-x^2= 1/{\rm GeV}^{2}$  (lower
panel) and $-x^2= 1/4\,{\rm GeV}^{2}$ (upper panel). Models for
pion DA are the same as used for the LO predictions, shown by
dotted curves, in Fig.~\ref{Fig-LO}.}
\end{figure}

Beyond LO, one may define effective partial waves,
\begin{equation}
\label{Def-F-eff} {\cal F}^{\rm eff}_n(\rho,x^2) = {\cal
F}_n(\rho,-x^2 \mu^2;\alpha_s(\mu))  + x^2
\frac{\phi^{(4)}_n(\mu)}{\phi_n(\mu)} {\cal F}_n(\rho),
\end{equation}
that additively combine perturbative and higher twist corrections
together. Here, the evolution of the twist-four coefficient is
governed by the difference in the LO anomalous dimensions of the
corresponding operators:
\begin{equation}
\frac{\phi^{(4)}_n(\mu)}{\phi_n(\mu)} =
\left(\frac{\alpha_s(\mu)}{\alpha_s(\mu_0)}\right)^\frac{32/9-
\gamma_n^{(0)}}{\beta_0}
\frac{\phi^{(4)}_n(\mu_0)}{\phi_n(\mu_0)},
\end{equation}
where $ \gamma_0^{(0)}=0$ and  $ \gamma_2^{(0)}=50/9$. Note that
these effective partial waves depend on the non-perturbative
quantities, which induces some model dependence.

As it is demonstrated in Fig.~\ref{Fig-corTW4}, twist-four
contributions are significant at $-x^2= 1/{\rm GeV}^{2}$ (lower
panel) and much less important at $-x^2=1/4{\rm GeV}^{2}$ (upper
panel). In particular, for  $-x^2= 1/{\rm GeV}^{2}$ the value at
$\rho=0$ decreases by about $16\%$ which is comparable to the NLO
perturbative correction, whereas for  $-x^2=1/4{\rm GeV}^{2}$ the
decrease is only by $\sim 3.5\%$ which is roughly one third of the
respective NLO effect. Note that the $\rho$ dependence of separate
partial waves is not affected by the twist-four contributions.
Their magnitude is changing, however: the second partial wave is
enhanced and the lowest one somewhat suppressed by the twist-four
corrections. These effects depend linearly on $x^2$ and are
additionally suppressed by evolution.

\begin{figure}[ht]
\begin{center}
\mbox{
\begin{picture}(250,220)(0,0)
\put(0,128){\rotatebox{90}{$T(\rho,x^2 = -1/4{\rm GeV}^{2} )$}}
\put(15,98){\insertfig{7}{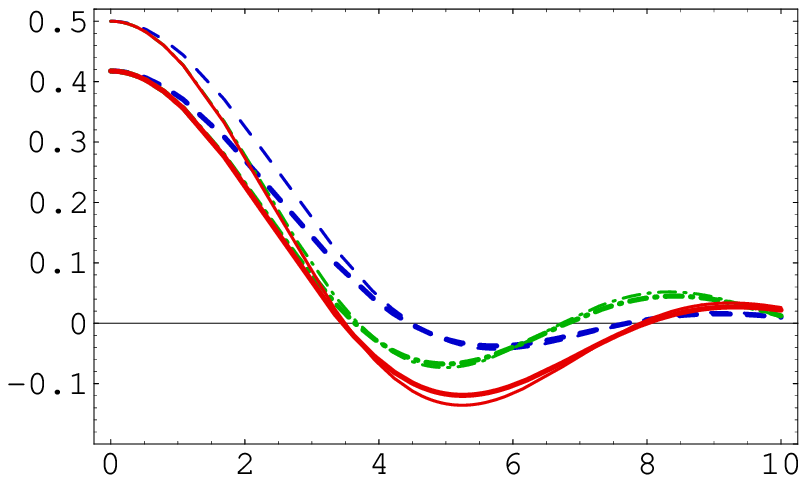}}
\put(0,20){\rotatebox{90}{$T(\rho,x^2 = -1/{\rm GeV}^{2} )$}}
\put(15,-10){\insertfig{7}{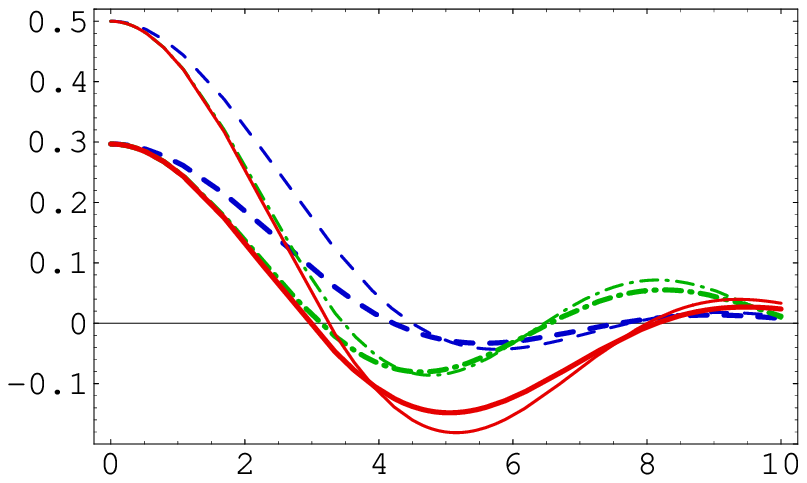}} \put(180,-15){$\rho=p\cdot
x$}
\end{picture}
}
\end{center}
\caption{ \label{Fig-Cor} \small Combined NNLO and twist-four
predictions for the correlation function (\ref{def:T}) in the
$\overline{\rm CS}$ scheme (thick curves) compared to the LO
results (thin curves). The factorization scale is set to the
quark-interquark separation, $\mu^2=-1/x^2$, which is $-x^2=1/{\rm
GeV}^{2}$ ($-x^2=1/4\,{\rm GeV}^{2}$) for the lower (upper) panel,
respectively. Models for pion DA are the same as in
Fig.~\ref{Fig-LO}.}
\end{figure}
Our final predictions combining the  perturbative effects to NNLO
accuracy in the $\overline{\rm CS}$ scheme and the twist-four
corrections are shown by thick curves in Fig.~\ref{Fig-Cor} and
compared to the corresponding leading-order leading twist
predictions (thin curves). {}For the separation $-x^2  =  1/4{\rm
GeV}^{2}$ the overall
 correction is rather small.
The normalization at $\rho=0$ is reduced by about $17\%$, whereas
the zeros and positions of extrema are essentially fixed. We
remind that there will be small flow of these points in the
$\overline{\rm MS}$ scheme, cf.~upper panel in
Fig.~\ref{Fig-MScorPT}. The overall corrections for $-x^2 = 1/{\rm
GeV}^{2}$ are much larger.

The conclusion is that the correlation function (\ref{def:T}) can
be calculated in QCD to a high accuracy. In particular for  $-x^2
\lesssim 1/4{\rm GeV}^{2} \sim (0.1~{\rm fm}^2$ the perturbation
theory works rather well and the higher--twist effects do not
exceed $3\%$.

The next question to address is whether the knowledge of the
correlation function (\ref{def:T}) can be used to constrain the
shape of the pion DA and, for example, allow for an accurate
determination of the few lowest Gegenbauer moments.

\section{Revealing the pion DA}
\label{Sec-RevDA}

Let us first consider the QCD prediction for the normalization of
the correlation function in the vicinity of $\rho=0$. Here only
the lowest partial wave contributes, while higher ones are
suppressed as
\begin{equation}
\label{Apr-F} |{\cal F}_n(\rho)| \le \frac{(n+1) (n+2) \sqrt{\pi
}}{4\Gamma(n+5/2)} \left|\frac{\rho}{2}\right|^n .
\end{equation}
Note the factorial suppression factor. For the favored value of
$\phi_2\sim 0.25$, we find for $\rho \simeq 1/4\, [\simeq 1/2]$
that the contamination of the second partial wave is  already
below three per mil [about 1\%], respectively. All higher partial
waves can certainly be omitted. We remind that a strong
suppression of higher partial waves occurs also in momentum space,
if both photons have a large virtuality \cite{Diehl:2001dg,
MelMuePas02}.  It follows that to three per mil [1\%] accuracy the
sum rule (\ref{SumRul}) can be extended for $\rho \le 1/4 [1/2]$:
\begin{eqnarray}
\label{SumRul-Ex} T(\rho\sim 0, x^2) &=&
\left[C_{0}(1,\alpha_s(-1/x^{2})) + \frac{8}{9} x^2
\delta^2_\pi\right] \frac{3}{4}{\cal F}_0(\rho)
\nonumber\\
&&+ O(\alpha_s x^2)+ O(x^4)\,,
\nonumber\\
\end{eqnarray}
where $(3/4){\cal F}_0(\rho) =1/2 -\rho^2/20 + O(\rho^4)$. We
emphasize again that in the $\overline{\rm CS}$ scheme the radiative
corrections to the lowest partial wave are absent at the
normalization point. We expect that the excitation of higher
partial waves due to the evolution yields a negligible $O(\rho^2
\alpha_s^2)$ effect, see also Eq.~(\ref{rotMSCS}) and  the
discussion of the $\overline{\rm MS}$ mixing. Hence, in this
scheme, the perturbative corrections can be borrowed from the
Bjorken sum rule, known to order $\alpha_s^3$ \cite{LarVer91}. For
shortness we display the two-loop result:
\begin{eqnarray}
&&C^{(2)}_{0}(-\mu^2 x^2, \alpha_s(\mu)) =
1-\frac{\alpha_s(\mu)}{\pi} - \frac{\alpha_s^2(\mu)}{\pi^2}
\\
&&\qquad  \times\left(\frac{55  - 4 n_f}{12}+\frac{\beta_0}{4}
\ln\left(-x^2 \mu^2 e^{2 \gamma_{\rm E}-1}\right)\right).
\nonumber
\end{eqnarray}
Note that due to the Fourier transform the argument of the
renormalization logs is decorated with a transcendent number $e^{2
\gamma_{\rm E}-1} \approx 1.166994\cdots$.

\begin{figure}[ht]
\begin{center}
\mbox{
\begin{picture}(250,120)(0,0)
\put(0,33){\rotatebox{90}{$T(\rho=1/4,x^2)$}}
\put(15,-10){\insertfig{7}{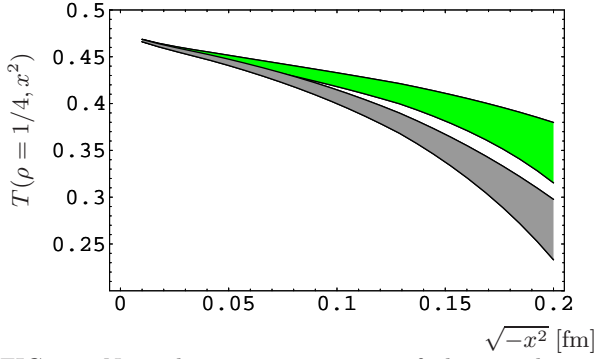}} \put(180,-20){$\sqrt{-x^2}\;
[{\rm fm}]$}
\end{picture}
}
\end{center}
\caption{ \label{Fig-QCD-pre} \small Normalization uncertainty of
the correlation function (\ref{def:T}) for $\rho=1/4$. The upper
error band  entirely arises by the uncertainty  of the running
coupling, while the lower one takes  also into account twist-four
contamination within the central value, given in
Eq.~(\ref{Tw4-par}).}
\end{figure}
To predict the normalization of the correlation function, we have
to specify the QCD coupling. The world average value of
$\alpha_s$, given at the $Z$-boson mass scale, reads to three loop
accuracy \cite{Bet00}
\begin{equation}
\label{Def-asNNLO} \alpha_s^{(3)}(\mu=91.18\, {\rm GeV}) = 0.1184
\pm 0.003\,.
\end{equation}
In the backward evolution to a lower scale we take into account
quark thresholds, treated in the standard way \cite{CheKuhSte00}.
{}For $\rho=0$ and the current separation  $-x^2 = 1/4{\rm
GeV}^{2} \sim (0.1~{\rm fm})^2$ we obtain, including three--loop
and higher--twist effects::
\begin{equation}
\label{Nor-T-rho0} T(0,-1/4 {\rm GeV}^2) = 0.430^{+0.008}_{-0.007}
- 0.018^{+0.006}_{-0.006},
\nonumber\\
\end{equation}
where the number of quarks is set to $n_f=4$. The predicted
twist-four correction is comparable to the error induced by the
uncertainty of $\alpha_s$, see Fig.~\ref{Fig-QCD-pre}, where the
dependence of $T(0,x^2)$, including the error bands, is displayed
as a function of $x^2$.

Next we discuss the access to the  Gegenbauer moments. We assume
that the correlation function is measured in a certain interval of
pion momenta alias an interval in $\rho$, $\rho_0 \le \rho \le
\rho_{\rm max}$. Since an overall normalization might be
difficult, we consider the ratio
\begin{eqnarray}
\label{Def-ratT}
{R}(\rho_0,\rho,x^2)=\frac{T(\rho,x^2)}{T(\rho_0,x^2)}\,,\quad
\rho_0< \rho\,,
\end{eqnarray}
which is normalized to one at $\rho=\rho_0$. Requiring that
$\rho_0 < 1$ allows to truncate the partial wave expansion in the
denominator at the second term $\sim\phi_2$. As the result,
${R}(\rho_0,\rho,x^2)$ is essentially a linear function of
$\phi_n$ with $n> 2$.

\begin{figure}[ht]
\begin{center}
\mbox{
\begin{picture}(250,120)(0,0)
\put(0,0){\rotatebox{90}{${\cal M}(1/2,\rho,x^2= -1/{\rm
GeV}^{2})$}} \put(15,-10){\insertfig{7}{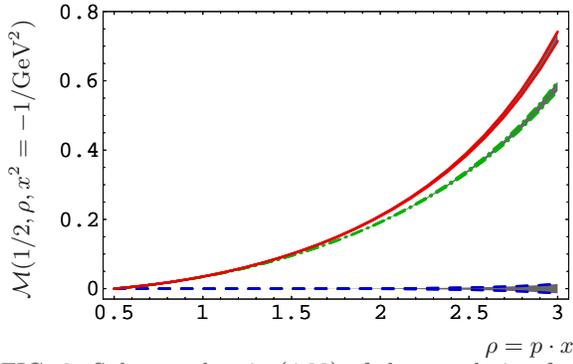}}
\put(180,-20){$\rho=p\cdot x$}
\end{picture}
}
\end{center}
\caption{ \label{Fig-Ext} \small Subtracted ratio (\ref{Def-M}) of
the correlation function (\ref{def:T}) calculated at LO as a
function of $\rho$ with $\rho_0=1/2$ and $-x^2= 1/{\rm GeV}^{2}$.
Models of the pion DA are same as in Fig.~\ref{Fig-LO}, where the
splitting of curves shows the possible (overestimated)
contribution of higher conformal partial waves with $n>4$. }
\end{figure}

The maximal accessible value of $\rho$, given by $\rho_{\rm max}$,
is limited by the pion momentum that is feasible in a lattice
calculation. We assume here that $\rho_{\rm max} < 3$, i.e., lower
than the first zero of the lowest conformal partial wave, cf.
Fig.~\ref{figharm}. Staying away from the zero allows one to
consider the subtracted ratio to access directly the second
Gegenbauer moment:
\begin{equation}
\label{Def-M} {\cal
M}(\rho_0,\rho,x^2)=1-\frac{R(\rho_0,\rho,x^2)}{{\cal
R}_0(\rho_0,\rho)}\,,\quad \rho_0 < \rho \lesssim 3\,,
\end{equation}
where ${\cal R}_n(\rho_0,\rho)= {\cal F}_n(\rho)/{\cal
F}_0(\rho_0)$. Since ${\cal R}_0(\rho_0,\rho)$ only decreases by
at most a factor three in the whole range $\rho \lesssim 3$, this
weight does not introduce any numerical instability. The conformal
spin expansion of ${\cal M}$ in terms of ratios of the effective
partial waves (\ref{Def-F-eff}) reads:
\begin{equation}
\label{Def-M-exp}
{\cal M}^{(4)}
\simeq \frac{\left[{\cal R}_2^{\rm eff}(\rho_0) -{\cal R}^{\rm
eff}_2(\rho)\right]\phi_2}{1+{\cal R}^{\rm eff}_2(\rho_0)\phi_2}-
\frac{{\cal R}_4^{\rm eff}(\rho)\phi_4}{1+{\cal R}^{\rm
eff}_2(\rho_0)\phi_2}\,,
\end{equation}
where
\begin{eqnarray}
{\cal R}^{\rm eff}_n(\rho,x^2) &=& \frac{{\cal F}^{\rm
eff}_n(\rho,x^2)}{{\cal F}_0^{\rm eff}(\rho,x^2)}\,.
\end{eqnarray}
Here we neglected all contributions in $\phi_6$, $\phi_8$, etc.
from the expansion in the numerator. The corresponding truncation
error is estimated to be  on the per mil [percentage] level for
$\rho < 2 [3]$, respectively. For example, for $\rho = 3$ we find
to LO accuracy:
\begin{equation}
{\cal M} - {\cal M}^{(4)}\sim 3\cdot 10^{-4} \phi_4 + 7\cdot
10^{-2} \phi_6 + 3\cdot 10^{-3} \phi_8  \,.
\end{equation}
Here the first term on the r.h.s.~comes from the truncation in the
dominator (assuming $\rho_0 =1/2$) and can safely be neglected,
same as the contribution of $\phi_8$. The remaining term in
$\phi_6$  effectively induces a small additive uncertainty in
determination of the first two Gegenbauer moments. For instance,
assuming values of  ${\cal M}^{(4)}$ are known for two different
values $\rho=2$ and  $\rho=3$, we find by solving the set of two
linear equations:
\begin{eqnarray}
\delta\phi_2 \approx  -0.01 \phi_6\,\quad \delta\phi_4 \approx
-0.13  \phi_6\,.
\end{eqnarray}
Assuming  $\phi_6\sim 0.1$ the errors induced by the truncation of
the partial wave expansion are in this version negligible for
$\phi_2$ and probably  on the one percent level for $\phi_4$. We
remind that convergency of the Gegenbauer expansion in the vicinity of
$u=0$ ($u=1$), which is required if the DA vanishes at the end points,
leads to an upper bound for the large--$n$ behavior of the
coefficients $|\phi_{n}| < {\rm const}/n^p$ with $p> 1$, so that
large values of $\phi_n$ with $n>4$ are increasingly unlikely.
Clearly, a fitting procedure including $\phi_6$ is also possible.
This is demonstrated in Fig.~\ref{Fig-Ext}, where the effect on
${\cal M}$ is shown for the $\phi_6$  varied in the interval
$-0.2\cdots 0.2$, which we consider as an overestimation.

\begin{figure}[ht]
\begin{center}
\mbox{
\begin{picture}(250,120)(0,0)
\put(0,0){\rotatebox{90}{${\cal M}(1/2,\rho,x^2= -1/4\,{\rm
GeV}^{2})$}} \put(15,-10){\insertfig{7}{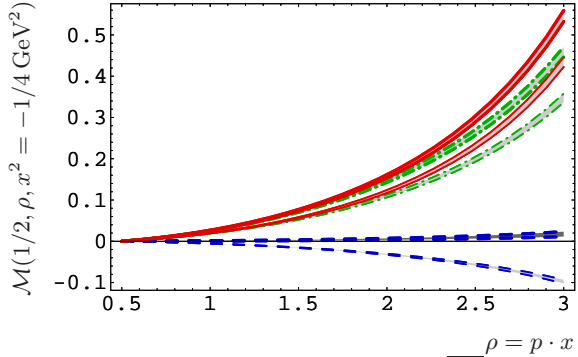}}
\put(180,-20){$\rho=p\cdot x$}
\end{picture}
}
\end{center}
\caption{ \label{Fig-Ext-R} \small Same as in Fig.~\ref{Fig-Ext}
to NLO for the $\overline{\rm MS}$  (thin) and $\overline{\rm CS}$
radiative (thick) schemes. The variation within the predicted
twist-four corrections are shown as error band. Here $-x^2=
1/4\,{\rm GeV}^{2}$ and the coupling is specified by the central
values (\ref{Def-alphas}). }
\end{figure}

Radiative corrections to the ratio (\ref{Def-ratT}) of the
correlation function are much milder as compared to their overall
normalization. A precision analysis requires a careful
specification of the running coupling. We take the world average
value (\ref{Def-asNNLO}) and calculate the corresponding value at
the scale $\mu=2\, {\rm GeV}$ by the tree loop evolution equation
within $n_f=4$:
\begin{eqnarray}
\label{Def-alphas} \alpha_s(\mu=2\, {\rm GeV})&=&
0.304^{\rm + 0.024}_{\rm - 0.021}.
\end{eqnarray}
We note that the resulting central value and errors of $\alpha_s$
at two loop accuracy are compatible with the given one; it is,
however, in the heart of perturbation theory that the
specification of $\alpha_s$ to LO accuracy goes hand in hand with
large uncertainties. As discussed in Sect.~\ref{Sec-PerCor}, the
asymptotic DA at NLO in the $\overline{\rm MS}$ differs from that
in the $\overline{\rm CS}$ scheme, mainly for the $n=2$ and $n=4$
moments, see Eq.~(\ref{AsyDA-MS}). This effect is  seen in
Fig.~\ref{Fig-Ext-R}, where the subtracted ratio $\mathcal{M}$
(\ref{Def-M-exp}) calculated in the $\overline{\rm MS}$ scheme
using the LO asymptotic DA (thin dashed curve) clearly deviates
from zero. This deviation is removed for the corrected asymptotic
DA (\ref{AsyDA-MS}), for which the $n=2$ moment is $\approx 0.04$.
For the same reason, the other two model predictions in the
$\overline{\rm MS}$ scheme (thin curves) are systematically
shifted downwards compared to the $\overline{\rm CS}$ (thick
curves). A small deviation from zero for the prediction
corresponding to the asymptotic DA in the $\overline{\rm CS}$
scheme (thick dashed curve) comes from the excitation of the $n=2$
partial wave  by a twist-four contribution. The shown narrow error
bands arise from the variation of the twist-four parameter in the
range $\delta_\pi^2(\mu = 2\,{\rm GeV}^2) = 0.09...0.19\, {\rm
GeV}^2.$

Since   perturbative corrections to the partial waves  in the
$\overline{\rm CS}$  scheme only induce a small modification of
the $\rho$ dependence in the considered range (absent for the
lowest one and on the one percent  or lower  level for higher
ones), the extraction of the first few Gegenbauer moments can be
considerably simplified. To this end, we define the effective
Gegenbauer moments:
\begin{equation}
\label{Def-Phi} \Phi_n^{\rm eff}=
\frac{C_n(\langle\langle\rho\rangle\rangle_n
|\alpha_s,-x^2\mu^2)\phi_n+ x^2
\phi_n^{(4)}}{C_0(\alpha_s,-x^2\mu^2)+ (8/9) x^2 \delta_\pi^2}\,,
\end{equation}
which depend on the average value of $\rho$ in the given range
($\langle\langle\rho\rangle\rangle_n\sim 2$ for  $\rho_{\rm
max}\sim 3$).

The subtracted ratio (\ref{Def-M-exp}) in the range  $\rho_0 \leq
\rho\leq \rho_{\rm max}$ can be written to a good accuracy in
terms of the effective moments as
\begin{equation}
\label{Def-M-exp-1}
{\cal M}^{(4)}
\simeq \frac{\left[{\cal R}_2(\rho_0) -{\cal
R}_2(\rho)\right]\Phi^{\rm eff}_2}{1+{\cal R}_2(\rho_0)\Phi^{\rm
eff}_2}- \frac{{\cal R}_4(\rho)\Phi^{\rm eff}_4}{1+{\cal
R}_2(\rho_0)\Phi^{\rm eff}_2}\,,
\end{equation}
where ${\cal R}_n(\rho) = {\cal F}_n(\rho)/{\cal F}_0(\rho)$ is
the ratio of the LO partial waves (\ref{Def-F-LO}).

The relation between the effective moments and the physical
Gegenbauer moments in the expansion of the pion DA includes
radiative and higher twist corrections. As an illustration, for
$\langle\langle\rho\rangle\rangle_n = 2.2$ and at $\mu = 2\, {\rm
GeV} $ we find
\begin{eqnarray}
\phi_2(2\, {\rm GeV}) &=& \left[0.961^{+0.003}_{-0.003} -
0.03^{+0.01}_{-0.01}\right] \Phi^{\rm eff}_2
\nonumber\\
&&{}+0.005^{+0.002}_{-0.002} +\cdots,
\\
\phi_4(2\, {\rm GeV}) &=& \left[0.902^{+0.008}_{-0.008} -
0.03^{+0.01}_{-0.01}\right] \Phi^{\rm eff}_4 +\cdots.
\nonumber
\end{eqnarray}
The first term in the square brackets corresponds to the
perturbative correction  while the second one arises from the
twist-four contribution. Note that the normalization change due to
radiative corrections is only on the level of 4\% for $\phi_2$ and
10\% for $\phi_4$, respectively, and is comparable with the higher
twist correction. In NNLO we find a slight increase of the
normalization $0.961 \to 0.982$ and $0.902 \to 0.951$ (for
$n_f=4$), whereas the error is drastically reduced. The residual
factorization and the renormalization scale dependence also prove
to be very small.

As the final step, the Gegenbauer moments in the $\overline{\rm
MS}$ scheme to NLO follow from
\begin{eqnarray}
\phi_2^{\overline{\rm MS}} &=&  \phi_2 + \frac{7}{18}
\frac{\alpha_s}{\pi} + O(\alpha_s^2)\,,
\\
\phi_4^{\overline{\rm MS}} &=&  \phi_4 +\frac{209}{810}
\frac{\alpha_s}{\pi}\phi_2 + \frac{11}{90} \frac{\alpha_s}{\pi}+
O(\alpha_s^2)\,. \nonumber
\end{eqnarray}

\section{Conclusion}
\label{Sec-Lat}

A typical setup for a lattice calculation of the correlation
function (\ref{lo:1}) would be to take the separation between the
two currents purely spacelike, $x^\mu \to \{0,\vec{x}\}$, and
integrate over the c.m.~position of the currents with a certain
three-dimensional momentum. Since the pion coupling to the source
is usually momentum dependent, the simplest strategy would
probably be to consider ratios of the correlation function
calculated for the same pion momentum but different $\vec{x}$ (and
different c.m.~momenta of the pair of currents), thus creating the
set of ``data points'' with correlated values of $x^2=
-|\vec{x}|^2$ and $\rho = \vec{p}\cdot\vec{x}$. In this way the
dependence on the pion coupling cancels out. The crucial condition
for such a calculation is to have a sufficient ``lever arm'' in
$\rho$. {} For a lattice spacing $a$ the momentum cannot be larger
than
$$ |\vec{p}| < a^{-1}\,, $$
which translates to the restriction on the maximal accessible value
of $\rho$
$$ \rho_{\rm max} < \frac{|\vec{x}|}{a}.$$
In other worlds, $\rho$ cannot be larger than (half of) the
separation between the currents in lattice units. Since, on the
other hand, perturbative treatment of the correlation function is
possible for $|\vec{x}|\le 0.2$~fm, this requirement translates to
a necessity to work on a fine lattice with $a < 0.03-0.05$ fm.
This presents a considerable challenge for an unquenched
calculation. The advantage of this formulation is, however, that
the problem of (nonperturbative) renormalization of lattice
operators is avoided altogether.

{}For a demonstration, we have chosen in this work to consider the
correlation function of two electromagnetic current in which case
the perturbative expansion is known to the NNLO accuracy. This
choice may be inconvenient for a lattice calculation because of
the epsilon-tensor appearing as a prefactor in (\ref{def:T}). This
problem can easily be avoided by considering a correlation
function of a vector and an axial-vector current, e.g.,
\begin{equation}
T'_{\mu\nu} = \langle 0| T\{\bar q(x) \gamma_\mu q(x) \bar q(-x)
\gamma_{\nu}\gamma_5 q(-x)\}\pi^0(p)\rangle\,.
\label{lo:1a}
\end{equation}

More importantly, the same strategy can be used to
extract information about the B-meson distribution
amplitude in the heavy quark limit from the lattice
measurement of the correlation function of the type
\begin{equation} \langle 0| T\{\bar h_v(x)
\gamma_\mu q(x) \bar q(-x) \Gamma q(-x)\} |B_v(p)\rangle\,,
\end{equation} where $v$ is the heavy quark velocity and $h_v$ an
effective heavy--quark field operator, which can be calculated in
terms of the B-meson DA using soft-collinear effective theory
\cite{DescotesGenon:2002mw,Bosch:2003fc,Beneke:2003pa}. The
standard method (calculation of the moments) is not applicable in
this case, since the existing definition of the B-meson DA relies
entirely on a perturbative factorization; the relation with the
Wilson operator expansion is lost unless an additional energy
cutoff is introduced, see \cite{Braun:2003wx,Lee:2005gza}.

Finally, nucleon distribution amplitudes
\cite{Chernyak:1983ej,Braun:2000kw} can be studied in the same
manner, from the correlation functions involving a local baryon
current.\\

\section*{Acknowledgements}
V.B. is grateful to IPPP for hospitality and financial support
during his stay at Durham University where this work was
finalized. This work has been partially supported by the
Verbundforschung (Hadrons and Nuclei) of the German Federal
Ministry for Education and Research (BMBF) (Contract 06 BO 103)
and by the EU Integrated Infrastructure Initiative Hadron Physics
Project under contract number RII3-CT-2004-506078.

\begin{appendix}

\renewcommand{\theequation}{\Alph{section}.\arabic{equation}}

\section{Perturbative expansion}
\label{AppPerQua}

\begin{widetext}
The radiative corrections in the position space might be obtained
from those in the momentum space by Fourier transform:
\begin{equation}
\label{FT-Mom2Pos} \frac{i\epsilon_{\mu\nu\rho\sigma} x^\rho
p^\sigma}{8\pi^2 x^4} T(p\cdot x, x^2) = \int\!
\frac{d^4q}{(2\pi)^4}\; e^{2i\,q\cdot x}\;
\frac{i\epsilon_{\mu\nu\rho\sigma} q^\rho p^\sigma}{q^2}
\widetilde{T}(\omega, -q^2)\,,
\end{equation}
where the hard scattering amplitude $\widetilde{T}(\omega, -q^2)$
is expressed by the asymmetry variable $\omega = -q\cdot p/q^2$
and $q= (q_1-q_2)/2$. Note that to LO accuracy the normalization
is chosen to be $\widetilde{T}(\omega=0)=1$. The correlation
function is given by the partial wave expansion:
\begin{equation}
\label{Def-T-MomSpa} \widetilde{T}(\omega, -q^2) = \sum_{n=0 \atop
{\rm even}}^\infty \widetilde{\cal F}_n(\omega,
-\mu^2/q^2;\alpha_s(\mu)) \phi_n(\mu^2)\,.
\end{equation}
In the hypothetical conformal limit the running coupling is
replaced by its non-trivial fixed point value, at which the
$\beta$-function vanishes. Then the predictive power of the conformal
operator product expansion can be used to obtain the partial
waves:
\begin{equation}
\label{Def-F-CS} \widetilde{\cal F}_n(\omega, -\mu^2/q^2|\gamma_n)
=  c_{n}(\alpha_s)  \frac{(n+1) (n+2) \sqrt{\pi}}{2^{n+2}
\Gamma(n+5/2)} \left(\frac{\mu^2}{-q^2}\right)^{\epsilon} \omega^n
{_2F_1}\!\left({(n+1+\epsilon)/2,(n+2+\epsilon)/2\atop
(n+5+\epsilon)/2}\Big|\omega^2\!\right)
\Bigg|_{\epsilon=\gamma_n/2}\,. \nonumber
\end{equation}
Here
\begin{equation}
\gamma_{n}(\alpha_s) = \frac{\alpha_s}{2\pi}  \gamma_n^{(0)}  +
\frac{\alpha^2_s}{(2\pi)^2}  \gamma_n^{(1)} + O(\alpha_s^3)\,,
\quad \gamma^{(0)}_n = C_F\left(4 S_1(n+1)-\frac{2}{(n+1)(n+2)}
-3\right)
\end{equation}
are the anomalous dimensions and
\begin{eqnarray}
\label{Def-WilCoeNL} \hspace{-0.5cm}
c_{n} =1 + \frac{\alpha_s}{2\pi} c_n^{(1)}  + \cdots
\,, \quad c_n^{(1)} = C_F \left(S^2_1(n+1) -S_2(n+1) +\frac{3}{2}
S_1(n+2)
 + \frac{3-2 S_1(n)}{2(n+1)(n+2)}  -\frac{9}{2}\right)
\end{eqnarray}
are the Wilson coefficients, normalized to one ($c_n^{(0)}\equiv
1$) at LO, of the polarized deeply inelastic structure function
$g_1$ for the flavor non-singlet sector. Both of them are
evaluated in the momentum space within the $\overline{\rm MS}$
scheme. The NLO anomalous dimensions can be found, e.g., in
Ref.~\cite{FloRosSac77}, and the few lowest one to NNLO in
Ref.~\cite{RetVer00}. The NNLO expressions (or numerical values
for the few lowest) Wilson--coefficients $c_{n}$  are obtained
from Ref.~\cite{ZijNee94}. All numbers, used in this paper, are
collected in Ref.~\cite{MelMuePas02}.

To transform the conformal predictions (\ref{Def-F-CS}) into the
position space, we first utilize the standard integral
representation for hypergeometric functions and in addition a
quadratic transformation:
\begin{eqnarray}
{_2F_1}\!\left({(n+1+\epsilon)/2,(n+2+\epsilon)/2\atop
(n+5+\epsilon)/2}\big|\omega^2\!\right)=
\frac{\Gamma(2n+4+2\epsilon)}{\Gamma(n+2+\epsilon)^2}
\int_0^1\!du\; \frac{ [(1-u)u]^{n+1+\epsilon} }{(1+\omega -2
\omega u)^{n+1+\epsilon}} \,.
\end{eqnarray}
Plugging Eq.~(\ref{Def-F-CS}) into Eq.~(\ref{Def-T-MomSpa}), and
interchanging $u$ and $q$ integration in Eq.~(\ref{FT-Mom2Pos}),
the Fourier transform can now be easily performed by means of:
\begin{eqnarray}
\label{FT} \int\! \frac{d^4q}{(2 \pi)^2}\, e^{2i x\cdot q}
\left(\frac{\mu^2}{-\bar{q}^2}\right)^{\epsilon}
\frac{\bar{q}_\rho}{\bar{q}^2} \left(\frac{p\cdot
\bar{q}}{-\bar{q}^2}\right)^n = \frac{1}{16  \pi^2} \frac{\Gamma
\left(2-\epsilon\right)}{\Gamma(1+n+\epsilon)} \frac{x_{\rho}
(-x^2  \mu^2)^{\epsilon}}{(-x^2)^2}(-i p\cdot x)^n e^{i(2u-1)
x\cdot p},
\end{eqnarray}
where $\bar q = q +(2u-1) p/2 $. Here $p^2$  and $p_\rho$ terms
are neglected to leading twist accuracy (the latter drops out
after contraction with the Levi-Civita tensor). Finally, the
$u$-integration leads to the partial waves (\ref{Def-ParWav-CS})
in position space  within the Wilson coefficients
(\ref{Def-ParWav-CS-coe}).

The conformal symmetry breaking due to the trace anomaly appears
the first time as $\alpha_s^2 \beta_0$ correction. It contains the
renormalization logs, needed to restore the renormalization group
invariance in the perturbative expansion with respect to a running
coupling $\alpha_s(\mu)$, and induces a mixing of Gegenbauer
moments, which depends on the factorization scheme. It is purely
conventional how we are dealing with this mixing. Appealing
possibilities are
\begin{enumerate}
\item[i.\phantom{ii}]
combining the conformal predictions (\ref{Def-F-CS}) with the
diagrammatical $\overline{\rm MS}$ result \cite{MelNicPas02},
\item[ii.\phantom{i}]
the partial waves  matches at the normalization point $-q^2=\mu^2$
the conformal predictions (\ref{Def-F-CS}),
\item[iii.] the Gegenbauer moments evolve autonomously.
\end{enumerate}
Certainly, the latter case is the most appealing ones, however, it
requires the full NNLO result for the trace anomaly induced term
(in an arbitrary scheme). Unfortunately, the corresponding piece
of the NNLO anomalous dimensions is unknown. Hence, we are only
able to follow the first two suggestions, where the mixing due to
evolution remains unknown \cite{MelMuePas02}. It turned out that
the numerical differences of the constant $\beta_0 \alpha_s^2/2$
proportional terms of these both schemes, are moderate. Moreover,
it is expected that the mixing effect due to the evolution is
tiny. We remind that in the $\overline{\rm MS}$ scheme such a
mixing under evolution already appears at NLO and leads in fact
only to small numerical effects.

Following  the second suggestion, we present here the NNLO
corrections in the so-called $\overline{\rm CS}$ scheme. To do so
we restore in the conformal predictions (\ref{Def-F-CS}) the scale
dependence of the coupling, i.e., $\alpha_s\to\alpha_s(\mu)$, and
the renormalization logs, which are governed by the
renormalization group equation
\begin{eqnarray}
\left[\mu\frac{\partial}{\partial \mu} +
\beta(\alpha_s)\frac{\partial}{\partial g}\right] \widetilde{\cal
F}_n(\omega, -\mu^2/q^2;\alpha_s(\mu))  = \gamma_n(\alpha_s(\mu))
\widetilde{\cal F}_n(\omega, -\mu^2/q^2;\alpha_s(\mu)) +
O(\alpha_s^3)\,,
\end{eqnarray}
approximated to NNLO.  The remaining freedom is fixed by the
requirement that at the normalization point $-q^2=\mu^2$ the
partial waves coincide with those in Eq.~(\ref{Def-F-CS}). A more
detailed discussion is given in Ref. \cite{MelMuePas02}. The so
found result,
\begin{equation}
\widetilde{\cal F}_n(\omega, -\mu^2/q^2;\alpha_s(\mu)) =
\widetilde{\cal F}_n(\omega, -\mu^2/q^2|\gamma_n(\alpha_s(\mu))) -
\frac{\beta_0}{2} \frac{\alpha_s^2(\mu)}{(2\pi)^2}
\Delta_{n}^{(2,\beta)}(\omega,-\mu^2/q^2) + O(\alpha_s^3)\,,
\end{equation}
is understood as a perturbative expansion up to NNLO accuracy,
where the addenda,
\begin{eqnarray}
\label{Def-Cbeta-MomSpa} \Delta_{n}^{(2,\beta)}(\omega,-\mu^2/q^2)
=  \ln(-\mu^2/q^2) \frac{\partial}{\partial (\alpha_s/2\pi)}
\left(1 + \frac{1}{2}\gamma_n(\alpha_s)\ln(-\mu^2/q^2)\!\right)
\widetilde{\cal F}_n(\omega,
-\mu^2/q^2|\gamma_n(\alpha_s))\Bigg|_{\alpha_s=0}\,,
\end{eqnarray}
restores the renormalization group invariance.

Again utilizing Eq.~(\ref{FT}) as expansion with respect to
$\epsilon$, the Fourier transform of the addenda
(\ref{Def-Cbeta-MomSpa}) is straightforward. Combining the
resulting expression with the perturbative expansion of the
conformal predictions (\ref{Def-ParWav-CS}) and
(\ref{Def-ParWav-CS-coe}) we find up to order $\alpha_s^2$:
\begin{equation}
\label{Def-PW-CS-NNLO} {\cal F}_n(\rho, -x^2 \mu^2;\alpha_s(\mu))=
\left[1 + \frac{\alpha_s(\mu)}{2\pi} C^{(1)}_{n}(\rho,-x^2\mu^2) +
\frac{\alpha_s^2(\mu)}{(2\pi)^2} C^{(2)}_{n}(\rho,-x^2\mu^2) +
O(\alpha_s^3)\right]\, {{\cal F}_n(\rho)}\,,
\end{equation}
where ${{\cal F}_n(\rho)}$ is the LO partial wave (\ref{Def-F-LO})
and
\begin{equation}
C_{n}^{(1)}=
 c_n^{(1)} + \frac{s_n^{(1)}(\rho,-x^2\mu^2)}{2} \gamma_n^{(0)}\,,
 \quad
C_{n}^{(2)}= \!c_n^{(2)} + \frac{s^{(1)}_n}{2}
\left(\gamma_n^{(1)}+c_n^{(1)} \gamma_n^{(0)}\right)+
\frac{s^{(2)}_n}{8} \left(\!\gamma_n^{(0)}\!\right)^2
-\frac{\beta_0}{2}C_{n}^{(2,\beta)} \,.
\end{equation}
The `shift' functions
\begin{eqnarray}
s_n^{(m)}(\rho,-x^2\mu^2)= \frac{d^m}{d\epsilon^m} (-\mu^2
x^2)^{\epsilon} \frac{\Gamma(2-\epsilon)
\Gamma(1+n)}{\Gamma(1+n+\epsilon)} \frac{{\cal
F}_n(\rho,-x^2\mu^2|2\epsilon)}{{\cal
F}_n(\rho)}\Bigg|_{\epsilon=0}
\end{eqnarray}
depend on the variables $-x^2\mu^2$ and $\rho$. To obtain the
renormalization group improved Wilson--coefficients we restore the
scale dependence of the coupling, i.e., $\alpha_s\to \alpha_s(\mu)
$, and take into account
\begin{equation}
C_{n}^{(2,\beta)} =  \left[ \left(\!\frac{\gamma_n^{(0)}}{4} {\rm
Ln}_n(-x^2 \mu^2)-C^{(1)}_{n}(\rho,-x^2\mu^2) \!\right){\rm
Ln}_n(-x^2 \mu^2) + \frac{3}{4}
\gamma_n^{(0)}\left(1-S_2(n)\right) \right],
\end{equation}
where ${\rm Ln}_n(-x^2 \mu^2)= \ln(-x^2 \mu^2)-S_{1}(n)+2
\gamma_{\rm E} -1$. We finally remark that the following
substitutions restores the result in momentum space:
\begin{eqnarray}
{{\cal F}_n(\rho)}\ &\Rightarrow&\ \widetilde{\cal F}_n(\omega)=
\frac{(n+1) (n+2) \sqrt{\pi}}{2^{n+2} \Gamma(n+5/2)} \omega^n
{_2F_1}\!\left({(n+1)/2,(n+2)/2\atop
(n+5)/2}\Big|\omega^2\!\right)\,,
\nonumber\\
s^{(m)}_n(\rho,-x^2\mu^2)  \ &\Rightarrow&\
s^{(m)}_n(\omega,-\mu^2/q^2) = \frac{d^m}{d\epsilon^m}
\left(\!\frac{\mu^2}{-q^2}\!\right)^{\epsilon}
\frac{{_2F_1}\!\left({(n+2\epsilon+1)/2,(n+2\epsilon+2)/2\atop
n+2\epsilon+5/2}\big|\omega^2\!\right)}{
{_2F_1}\!\left({(n+1)/2,(n+2)/2\atop
n+5/2}\big|\omega^2\!\right)}\Bigg|_{\epsilon=0},
\\
C_{n}^{(2,\beta)}(\rho,-\mu^2 x^2)  \ &\Rightarrow&\
C_{n}^{(2,\beta)}(\omega,-\mu^2/q^2) =
\left(\!C^{(1)}_{n}(\omega,-\mu^2/q^2) + \frac{1}{2}\gamma_n^{(0)}
\ln(-\mu^2/q^2)\!\right)\ln(-\mu^2/q^2)\,. \nonumber
\end{eqnarray}

The NLO corrections to the correlation function in the
$\overline{\rm MS}$ scheme might be  obtained by Fourier transform
from the known result in momentum space
\cite{AguCha81,Bra83,KadMikRad86}. Representing the
hard--scattering part as a convolution of the LO hard-scattering
part within some kernels, see Ref.~\cite{Mue97a}, the
transformation can be straightforwardly performed by utilizing
Eq.~(\ref{FT}) for $n=0$. Alternatively, we can simply rotate the
partial waves (\ref{Def-PW-CS-NNLO}) in the $\overline{\rm CS}$
scheme, expanded up to NLO. Note that the trace anomaly does not
enter in this approximation. The non--vanishing entries of the
rotation matrix, used in Eq.~(\ref{rotMSCS}),
\begin{eqnarray}
\label{Def-B} B^{(1)}_{nm}= \frac{2(2n\!+\!3) C_F }{
(n\!+\!1)(n\!+\!2)} \left[
\frac{(m\!+\!1)(m\!+\!2)}{(n\!-\!m)(n\!+\!m\!+\!3)} \left(2
A_{nm}-\frac{ \gamma^{(0)}_m}{2 C_F}
\right)+A_{nm}-S_1(n+1)\right],
\end{eqnarray}
appear for $n>m$ and $n-m$-even, where
\begin{equation}
A_{nm} = S_1((n+m+2)/2)-S_1((n-m-2)/2)+2 S_1(n-m-1) - S_1(n+1)\,.
\end{equation}

\end{widetext}

\end{appendix}


\end{document}